\documentclass[sigconf]{acmart}

\usepackage{colortbl}
\usepackage{makecell}
\usepackage{multirow}

\acmConference[MSR 2026]{MSR '26: Proceedings of the 23rd International Conference on Mining Software Repositories}{April 2026}{Rio de Janeiro, Brazil}

\AtBeginDocument{
  }

\copyrightyear{2026}
\acmYear{2026}
\setcopyright{cc}
\setcctype{by}
\acmConference[MSR '26]{23rd International Conference on Mining Software Repositories}{April 13--14, 2026}{Rio de Janeiro, Brazil}
\acmBooktitle{23rd International Conference on Mining Software Repositories (MSR '26), April 13--14, 2026, Rio de Janeiro, Brazil}
\acmPrice{}
\acmDOI{10.1145/3793302.3793622}
\acmISBN{979-8-4007-2474-9/2026/04}

\begin{document}
\title{More Code, Less Reuse: Investigating Code Quality and Reviewer Sentiment towards AI-generated Pull Requests}

\author{Haoming Huang}
\authornote{Equal contribution}
\email{haoming@se.comp.isct.ac.jp}
\affiliation{
  \institution{Institute of Science Tokyo}
  \city{Tokyo}
  \country{Japan}}

\author{Pongchai Jaisri}
\email{jaisri.pongchai.js3@naist.ac.jp}
\authornotemark[1]
\affiliation{
  \institution{Nara Institute of Science and Technology}
  \city{Nara}
  \country{Japan}}

\author{Shota Shimizu}
\email{is0618vf@ed.ritsumei.ac.jp}
\affiliation{
  \institution{Ritsumeikan University}
  \city{Ibaraki}
  \country{Japan}}

\author{Lingfeng Chen}
\email{lingfeng@posl.ait.kyushu-u.ac.jp}
\affiliation{
  \institution{Kyushu University}
  \city{Fukuoka}
  \country{Japan}}

\author{Sota Nakashima}
\email{nakashima@posl.ait.kyushu-u.ac.jp}
\affiliation{
  \institution{Kyushu University}
  \city{Fukuoka}
  \country{Japan}}

\author{Gema Rodríguez-Pérez}
\email{gema.rodriguezperez@ubc.ca}
\affiliation{
  \institution{University of British Columbia}
  \city{Kelowna}
  \country{Canada}}

\newcommand{\rqone}{How do AI-generated code differ from human-generated code in terms of code quality?}
\newcommand{\rqtwo}{How does reviewer sentiment differ between AI-generated and human-written PRs}

\begin{abstract}

Large Language Model (LLM) Agents are advancing quickly, with the increasing leveraging of LLM Agents to assist in development tasks such as code generation.
While LLM Agents accelerate code generation, studies indicate they may introduce adverse effects on development.
However, existing metrics solely measure pass rates, failing to reflect impacts on long-term maintainability and readability, and failing to capture human intuitive evaluations of PR.
To increase the comprehensiveness of this problem, we investigate and evaluate the characteristics of LLM to know the pull requests' characteristics beyond the pass rate.
We observe the code quality and maintainability within PRs based on code metrics to evaluate objective characteristics and developers' reactions to the pull requests from both humans and LLM's generation.
Evaluation results indicate that LLM Agents frequently disregard code reuse opportunities, resulting in higher levels of redundancy compared to human developers.
In contrast to the quality issues, our emotions analysis reveals that reviewers tend to express more neutral or positive emotions towards AI-generated contributions than human ones.
This disconnect suggests that the surface-level plausibility of AI code masks redundancy, leading to the silent accumulation of technical debt in real-world development environments. Our research provides insights for improving human-AI collaboration.

\end{abstract}


\maketitle

\section{Introduction}
\label{intro}

The software engineering (SE) landscape is undergoing a paradigm shift towards SE 3.0, where Large Language Model (LLM) agents are evolving from passive assistants to autonomous teammates \cite{liRiseAITeammates2025}. 
The AIDev dataset \cite{liRiseAITeammates2025} captures the emergence of pull requests (PRs) submitted autonomously by AI Agents (Agentic-PRs). 
While LLM Agents accelerate code generation, researches~\cite{beckerMeasuringImpactEarly20252025, perryUsersWriteMore2023} indicate they may introduce adverse effects on development. Existing benchmark (like SWE-Bench~\cite{jimenezSWEbenchCanLanguage2023}) solely measure pass rates, failing to reflect impacts on long-term maintainability and readability. While code redundancy or code clone is a well-known issue in SE \cite{roySurveySoftwareClone2007}, evaluating it in the context of AI agents remains unexplored.

A concern with AI-generated code is code clone~\cite{fowlerRefactoringImprovingDesign2018}, a.k.a code redundancy. In SE, code redundancy is a known bad smell that increases maintenance effort~\cite{fowlerRefactoringImprovingDesign2018}. 
Due to LLM’s probabilistic model-based nature~\cite{bengioNeuralProbabilisticLanguage2003, shumailovCurseRecursionTraining2024}, we believe generative AI models tend to produce \textit{Type-4} semantic clones~\cite{roySurveySoftwareClone2007, shumailovAIModelsCollapse2024}, unlike human developers who often copy and paste code (creating \textit{Type-1} or \textit{Type-2} clones~\cite{juergensCodeClonesMatter2009}). 
These clones implement the same logic as existing functions but use different syntax or variable names. This behavior violates the SE principle of reuse. 
However, existing tools are not effective at detecting these \textit{Type-4} semantic clones (see \autoref{related}).

To address this problem, we propose a comprehensive evaluation framework. We assess Agentic-PRs from two perspectives: internal code quality and external reviewer perception. For internal code quality, we measure traditional metrics including \textit{Lines of Code (LOC)} and \textit{Cyclomatic Complexity}~\cite{mccabeComplexityMeasure1976}. Furthermore, we propose a new metric called the \textit{Max Redundancy Score (MRS)} to measure semantic redundancy. We use code emebdding model to build an approach, to detect \textit{Type-4} semantic clones. For external reviewer perception, we investigate how human reviewers react to Agentic-PRs. We conduct a sentiment analysis on review comments to see if reviewers identify the quality issues in AI-generated code.

We analyze the AIDev dataset~\cite{liRiseAITeammates2025} to answer the following:
\begin{itemize}
    \item \textbf{RQ1:} \rqone
    \item \textbf{RQ2:} \rqtwo
\end{itemize}
This study provides the first empirical evidence of the disconnect between AI-generated code redundancy and human reviewer sentiment. By highlighting the risk of silent technical debt, we offer insights for improving human-AI collaboration.

\section{Methodology}
\label{method}

\subsection{Dataset}
\label{dataset}
We utilize the AIDev dataset~\cite{liRiseAITeammates2025} for our empirical study. To ensure the analysis focuses on mature projects, we filter the dataset to include only Python repositories with more than 500 stars, following the popularity standard mentioned in AIDev~\cite{liRiseAITeammates2025}. 
We separate the data into human-generated PRs (Human-PRs) and agent-generated PRs (Agentic-PRs). 
To balance the depth of semantic analysis with computational feasibility, we employ a two-level data strategy:

\begin{itemize}
    \item \textbf{Dataset A} (Full Python Dataset): This dataset comprises 3,858 PRs across all filtered Python repositories. We use this broad dataset to analyze traditional code metrics (RQ1) and reviewer sentiment (RQ2).
    \item \textbf{Dataset B} (Core Case Study): This dataset consists of 617 PRs from the crewAI repository, the repository with the most PRs in AIDev. As the most active repository in AIDev with high agent activity, it serves as a representative case for our computationally intensive redundancy analysis (RQ1).
\end{itemize}

\subsection{RQ1: \rqone}
This question compares the code quality between handcraft changes and AI-generated changes. We employ a multi-dimensional evaluation strategy including both traditional static metrics and a novel code redundancy analysis.

\begin{figure}[ht]
    \centering
    \includegraphics[width=\linewidth]{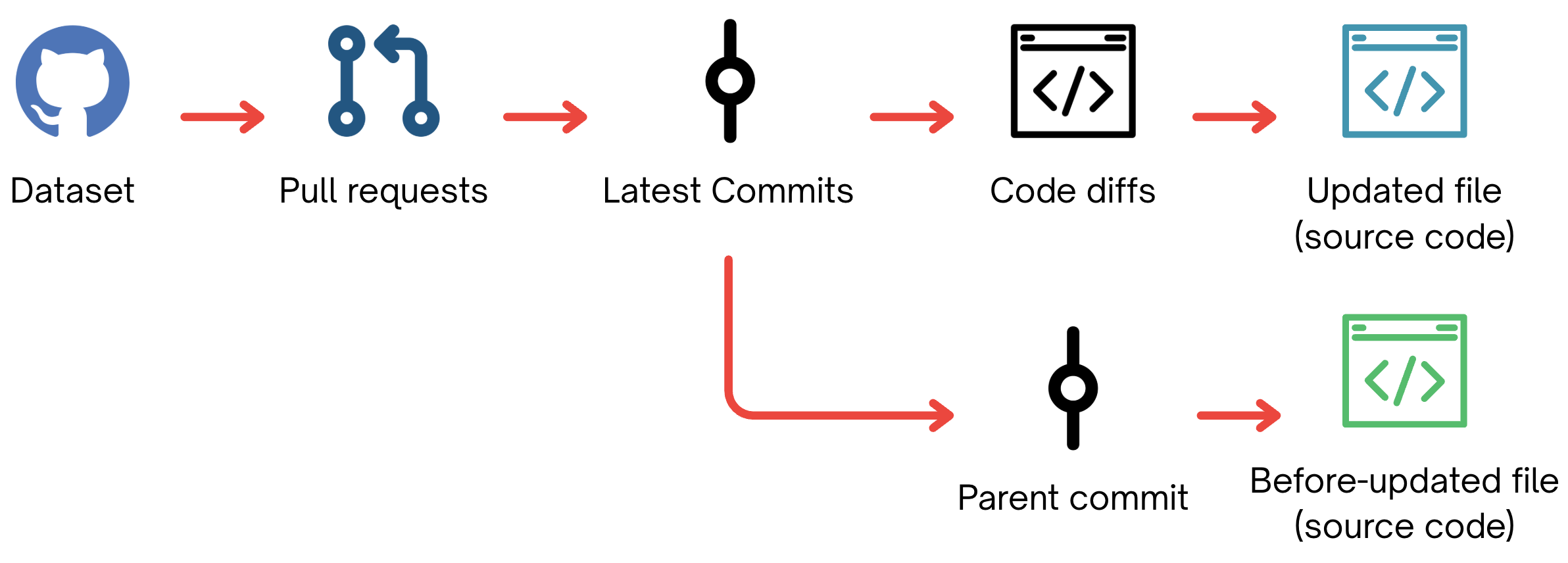}
    \caption{Overview of collecting changes from pull requests}
    \label{fig:rqone_methodology}
    \Description[Overview of collecting changes from pull requests]{This is a flowchart that starts from the Dataset, extracts the Pull Request, then extracts the final commit, and finally extracts the files before and after the update.}
\end{figure}

As shown in~\autoref{fig:rqone_methodology}, We extracted pairs of source files representing the pre- and post-update states by identifying the modified files in each PR relative to their parent commits.

\subsubsection{Traditional Metrics}

We calculated \textit{Cyclomatic Complexity (CC)} and raw metrics (\textit{LOC}, \textit{multi-lines-string}, \textit{blank lines}) using Radon\footnote{\url{https://pypi.org/project/radon/}} for each file pair to compare human and AI contributions.

\subsubsection{Redundancy Score} \label{sec:MRS}

Due to the probabilistic nature of LLM models~\cite{shumailovCurseRecursionTraining2024,bengioNeuralProbabilisticLanguage2003}, they are more prone to generating \textit{Type-4} Clones, which reduce software maintainability. While LLMs can also produce \textit{Type-1} and \textit{Type-2} clones, existing tools that rely on literal keyword matching remain ineffective at identifying the more prevalent \textit{Type-4} Clones.

To bridge this gap, we introduce a Snapshot-based Semantic Differential Analysis pipeline. 
We extract all existing functions from the snapshot ($F_{base}$) and generate semantic embeddings using CodeSage-Large \cite{zhangCodeRepresentationLearning2024}, a model optimized for code representation. 
This allows us to capture functional equivalence beyond syntactic similarity. 
A naive comparison of added code yields false positives, as moving or renaming a function appears as an addition in Git diffs. 
To isolate true redundancy, we integrate PyRef \cite{atwiPYREF21stIEEE2021}, a refactoring detection tool, to identify and exclude \textit{Move Method} and \textit{Rename Method} instances from the set of new functions ($F_{new}$). 
We define the redundancy of a PR not by its average similarity, but by its worst offender. 
In maintenance, a single instance of severe duplication (e.g., rewriting a complex core utility) creates a single point of failure for future updates \cite{juergensCodeClonesMatter2009}. 
Thus, we calculate the \textit{Max Redundancy Score (MRS)} by \autoref{eq:MRS};

\begin{equation} \label{eq:MRS}
\text{MRS}(P_i) = \max_{f \in F_{new}^{(i)}} \left( \max_{g \in F_{base}^{(i)}} \cos(\mathbf{v}(f), \mathbf{v}(g)) \right)
\end{equation}

where $\mathbf{v}(\cdot)$ denotes the embedding vector. We finally report the \textit{Average Max Redundancy (AMR)} to compare human and AI groups.

\subsection{RQ 2: \rqtwo}
To understand how human developers perceive code generated by AI agents, we analyze the sentiments expressed in code review comments. Specifically, we compare the emotional reactions in Agentic-PRs versus Human-PRs using Dataset A (the full Python dataset) for this analysis.

To obtain the sentiments, we use state-of-the-art sentiment analysis model, Emotion English DistilRoBERTa-base~\cite{hartmann2022emotionenglish}. This model classifies emotions to Ekman's six basic emotions (anger, disgust, fear, joy, sadness, surprise) and neutral. Since the model has a maximum token length of 512, we exclude the data which was more than maximum token lengths.

To compare the emotional reactions in Agentic-PRs versus Human-PRs, we analyze in the following process. First, we used code review comments from $pr\_comments$, $pr\_review\_comments\_v2$, $pr\_reviews$ from AIdev~\cite{liRiseAITeammates2025}.
To focus on developer sentiments, we exclude comments made by bots.
Second, we filter Dataset A to identify the PRs made by Agents and Humans. 
Next, we classify the sentiments of comments made by each PR using Emotion English DistilRoBERTa-base. 
We calculate the average sentiment scores per PR to obtain how developers perceived within each PR. 
This process was applied separately to Agentic-PRs and Human-PRs.

Additionally, we identify the PRs with the highest sentiment scores for each emotion in both Agentic-PRs and Human-PRs. 
Analyzing these PRs provides further insights into how the two groups differ in terms of sentiments expressed during code reviews.

\section{Results}
\label{results}

\subsection{Results of RQ1}
\label{rq1result}

\paragraph{\textbf{Raw Code Metrics:}}

As shown in \autoref{tab:locresult}, while code additions vary slightly between groups, human developers remove significantly ($p < 0.001$) more \textit{multi-lines-string} (26.26) than AI agents (8.47), alongside a higher average \textit{LOC} removal (25.51 vs. 16.60).

\begin{table}[htbp]
    \caption{Average Addition Lines Differences by Metric.}
    \label{tab:locresult}
    \small
    \centering
    \setlength{\tabcolsep}{5pt} 
    \renewcommand{\arraystretch}{0.95}
    
    \begin{tabular}{l|l | r|c|r}
        \toprule
        \toprule
        \multicolumn{1}{c}{\makecell{Change Types}} & PRs' types & \multicolumn{1}{l}{~~ LOC} & \multicolumn{1}{l}{\makecell{Multi- \\ Lines}} & \multicolumn{1}{c}{Blank Lines} \\
        \midrule
        Addition & Human    & 23.01\phantom{\textsuperscript{***}} & 15.79\phantom{\textsuperscript{***}} & 6.16\phantom{\textsuperscript{*****}} \\
                 & Agent    & 23.78\phantom{\textsuperscript{***}} & 16.48\phantom{\textsuperscript{***}} & 6.02\phantom{\textsuperscript{*****}} \\
                 & $\Delta$ Diff     & \cellcolor{red!20}-0.77\textsuperscript{***} & \cellcolor{red!20}-0.69\phantom{\textsuperscript{***}} & \cellcolor{green!20}0.14\textsuperscript{*}\phantom{\textsuperscript{****}} \\
        \midrule
        Removal  & Human    & 25.51\phantom{\textsuperscript{***}} & 26.26\phantom{\textsuperscript{***}} & 6.99\phantom{\textsuperscript{*****}} \\
                 & Agent    & 16.60\phantom{\textsuperscript{***}} & 8.47\phantom{\textsuperscript{***}} & 5.05\phantom{\textsuperscript{*****}} \\
                 & $\Delta$ Diff      & \cellcolor{green!20}8.91\phantom{\textsuperscript{***}}  & \cellcolor{green!20}17.79\textsuperscript{***} & \cellcolor{green!20}1.94\phantom{\textsuperscript{*****}} \\
        \bottomrule
        \bottomrule
        \multicolumn{5}{l}{\footnotesize{*~Significant ($p < 0.05$) \quad \quad \quad \quad \quad \quad \quad***~ Highly Significant ($p < 0.001$)}}

    \end{tabular}
    \Description[Comparison of average code metrics between Human and Agent PRs]{A table comparing average line counts for Addition and Removal changes across three metrics: LOC, Multi-Line Strings, and Blank Lines. The table is divided into two main sections: Addition and Removal. For each section, it lists values for Human, Agent, and the difference (delta). Significant findings include: 1) For Addition LOC, a very significant negative difference of -0.77 (p < 0.001), highlighted in red; 2) For Addition Blank lines, a significant positive difference of 0.14 (p < 0.05), highlighted in green; 3) For Removal Multi-Line String, a very significant positive difference of 17.79 (p < 0.001), highlighted in green, indicating humans remove significantly more comments than agents.}
\end{table}

\paragraph{\textbf{Cyclomatic Complexity (CC):}}
As shown in \autoref{tab:complexityresult}, the computation of the CC through all pairs, most of code changes or 18,968 pairs (85.02\%) have zero complexity score. The result shows that only 3,333 from 22,311 pairs (14.94\%) have scored more than 0 which include 1,343 pairs (6.02\%) are low risk code removal pairs, 1,979 pairs (8.87\%) are low risk code addition pairs and 11 pairs (0.05\%) are high risk code addition pairs.

\begin{table}[htbp]
    \centering
    \caption{Distribution of Complexity Score of Code Changes}
    \label{tab:complexityresult}
    \Description[Distribution of complexity scores for Human and Agent code changes]{Distribution of complexity scores for Human and Agent code changes}

    \begin{tabular}{l | c|c|c}
        \toprule
        \toprule
        \multirow{3}{*}{PRs' types} & \multicolumn{3}{c}{Complexity Score (CC)} \\
        \cmidrule(l){2-4}
        & 0 & 0 < |CC| < 10 & 10+ \\
        & No Risk & Low Risk & High Risk \\
        \midrule
        Human & 8,643  & 1,463  & 6 \\
        Agent & 10,325 & 1,859  & 5 \\
        \midrule
        All (PRs) & 18,968 & 3,322 & 11 \\
        \bottomrule
        \bottomrule
    \end{tabular}
\end{table}

\begin{figure}[htbp]
    \centering
    \includegraphics[width=0.94\linewidth]{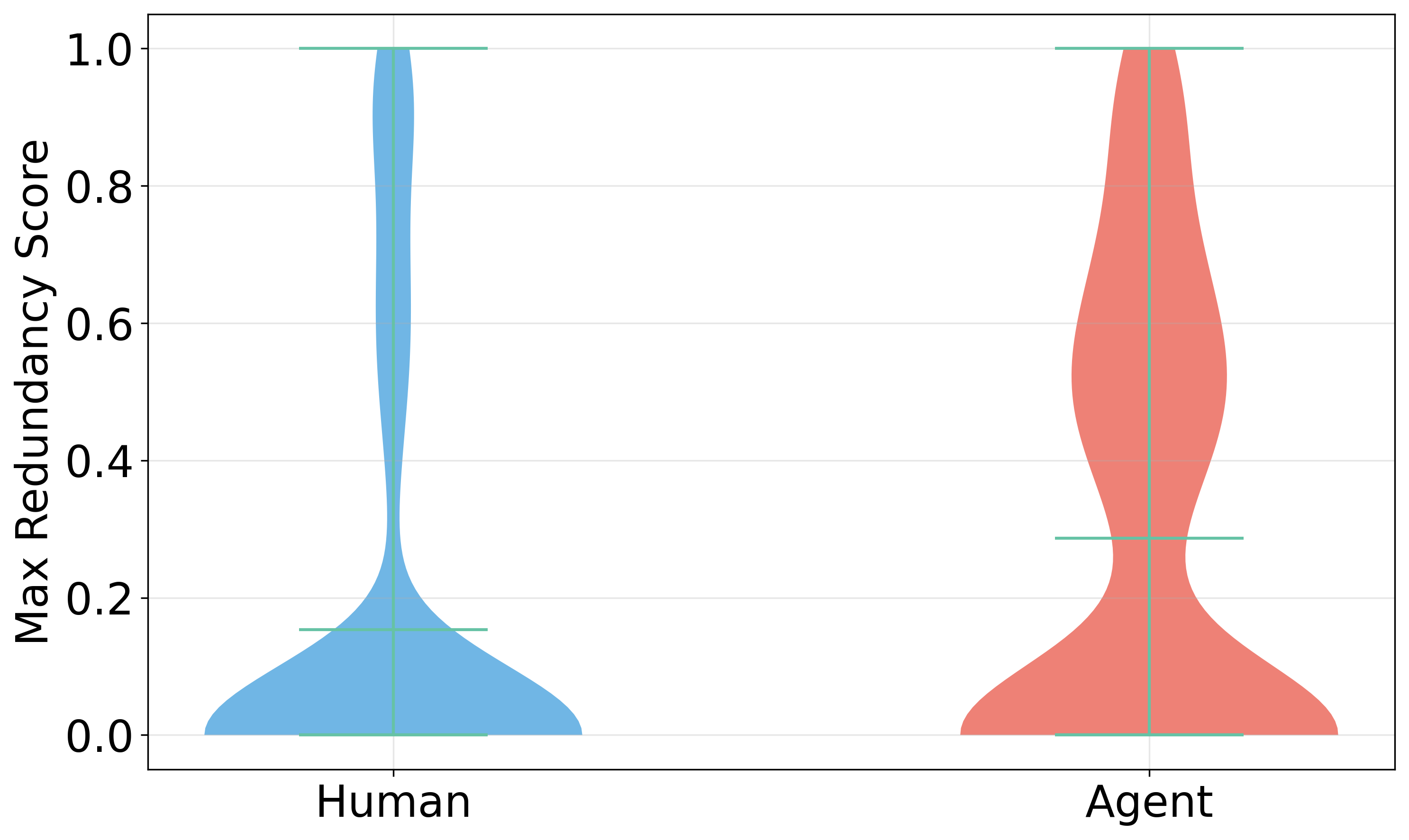}
    \caption{MRS Distribution Between Humans and AI}
    \vspace{-0.2cm}
    \label{fig:rq1_redundancy}
    \Description[MRS Distribution Between Humans and AI]{This is a comparison chart of MRS distributions. The blue distribution on the left represents human MRS, while the red distribution on the right represents AI agent MRS. In the region of low MRS (below 0.2), human and AI MRS are similar. In the intermediate MRS range (between 0.3 and 0.8), AI demonstrates greater redundancy.}
\end{figure}

\paragraph{\textbf{Redundancy:}}
As shown in \autoref{fig:rq1_redundancy}, Agentic-PRs demonstrate a significantly higher level of redundancy. 
For low \textit{Max Redundancy Score} (\textit{MRS}, see \autoref{sec:MRS}) sections, the distributions of AI agents and humans are largely similar. 
However, for medium \textit{MRS} sections (0.3–0.8), AI agents exhibit higher \textit{AMR}. 
We infer that both humans and AI agents produce some similar code (e.g., API calls or common patterns). 
However, LLM-generated code significantly exceeds human-written code in higher AMR ranges. 
Quantitatively, the \textit{Average Max Redundancy (AMR)} for AI agents is 0.2867, compared to only \textit{0.1532} for humans, representing a nearly \textit{1.87x increase}. Mann-Whitney test confirms that this difference is statistically significant ($p < 0.001$).

\vspace{0.5em}
\noindent
\fbox{
    \begin{minipage}{0.96\linewidth}
        \textbf{Answer to RQ1:} While traditional metrics show minimal differences between agentic-PRs and human-PRs, redundancy metric analysis shows code in \textit{agentic-PRs contain significantly more redundancy} ($p < 0.001$).
    \end{minipage}
}

\subsection{Results of RQ2}

\begin{figure*}[ht]
    \centering
    \includegraphics[width=0.87\linewidth]{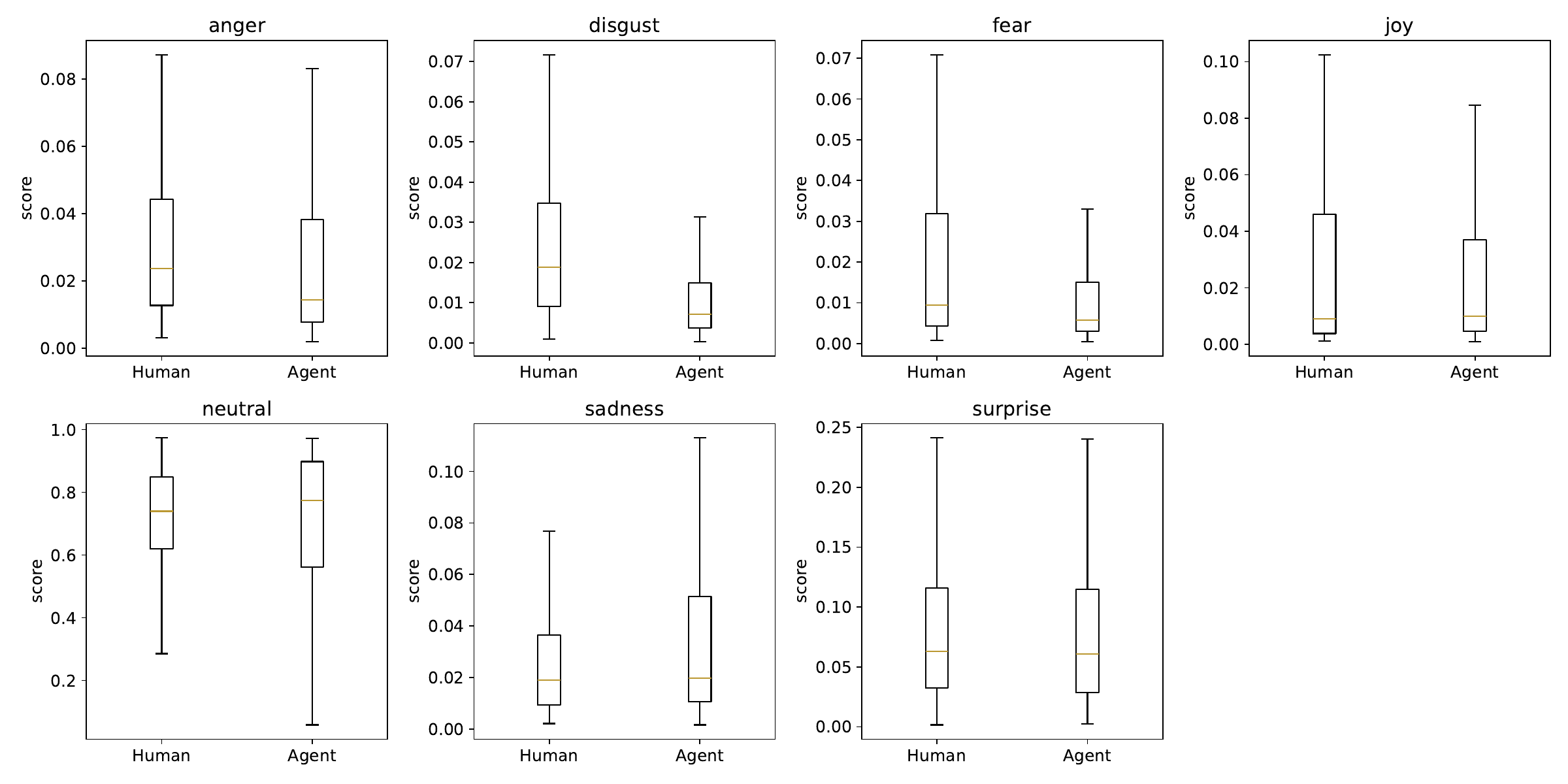}
    \caption{The reviewer sentiment score at Human-PRs and Agentic-PRs}
    \label{fig:boxplot_RQ2}
    \Description[The reviewer sentiment score at Human-PRs and Agentic-PRs]{This is a box plot comparing reviewer sentiment scores. It contains seven subplots, each illustrating the distribution differences between human reviewers and Agent PR in expressing seven emotions: anger, disgust, fear, joy, neutral, sadness, and surprise.}
\end{figure*}

The result of RQ2 is shown in \autoref{fig:boxplot_RQ2}.
The most frequent sentiment is neutral, meaning that reviewers mostly review PRs respectfully for both Human-PRs and Agentic-PRs. 
However, comparing the differences between each sentiment, Agentic-PRs had a higher proportion of neutral, joy, and sadness, while Human-PRs had a higher proportion of disgust, anger, fear, and surprise.

\vspace{0.5em}
\noindent
\fbox{
    \begin{minipage}{0.96\linewidth}
        \textbf{Answer to RQ2:} Our analysis shows that reviewers tend to express more \textit{neutral or positive} emotions towards AI-generated contributions than human ones. 
    \end{minipage}
}

\section{Discussion}
\label{discussion}

\subsection{Disconnect between Quality and Sentiment}

Despite the similarity in traditional metrics, we found a more critical issue on AI-generated PR. Our redundancy analysis (RQ1) shows that AI agents introduce significantly more redundancy than humans. AMR for Agentic-PRs is nearly double that of Human-PRs. This confirms that AI agents tend to generate redundant code instead of reusing existing code.

However, RQ2 shows that reviewers do not react negatively to this issue. In fact, reviewers express fewer negative emotions (such as anger or disgust) towards Agentic-PRs compared to Human-PRs. This counter-intuitive disconnect may stem from the training objectives of current LLMs. Models aligned using Reinforcement Learning from Human Feedback\cite{ouyangTrainingLanguageModels2022} are optimized to maximize human preference and perceived helpfulness~ \cite{casperOpenProblemsFundamental2023}. Existing studies observed that LLMs tend to prioritize sycophancy over maximizing the accuracy of truthful statements or code quality when responding\cite{ouyangTrainingLanguageModels2022, sharmaUnderstandingSycophancyLanguage2023, liDissectingHumanLLM2024, perryUsersWriteMore2023, weiSimpleSyntheticData2024, fanousSycEvalEvaluatingLLM2025}. 
Consequently, AI agents tend to produce code that appears superficially correct and helpful.

This disconnect suggests a reviewer's blind spot. AI-generated code is syntactically correct and often passes tests. Because LLM agents generate code based on probability, the output looks plausible on the surface. As a result, reviewers may lower their vigilance. They focus on whether the code works (pass rate), but they fail to check if the code duplicates existing logic in the repository.

\subsection{The Risk of Silent Technical Debt}
This situation leads to the accumulation of silent technical debt. When an AI agent introduces a Type-4 clone, it creates a duplicate maintenance point. Previous studies have demonstrated that inconsistent updates to duplicated code are a primary source of software defects \cite{juergensCodeClonesMatter2009}. If a bug exists in the logic, a developer might fix the original utility function but miss the redundant copy created by the AI.
Because reviewers are reacting positively (or neutrally) to these PRs, this debt accumulates unnoticed. Consequently, the codebase becomes bloated with redundant logic, which decreases long-term readability and increases future maintenance effort \cite{roySurveySoftwareClone2007}.

\subsection{Implications}

\textbf{For Agent Builders:} Researchers and tool developers need to address the issue of code redundancy. Our results show that current agents often ignore existing code. Therefore, builders should evaluate agents not just on pass rates, but also on code reuse metrics. 
\textbf{For Software Developers:} Although AI agents increase coding speed in the short term, developers must be aware of the impact on long-term maintainability. The plausible code from AI may contain hidden redundancy. Reviewers should actively check for duplicated logic during code reviews. Ignoring this now may lead to higher review costs and maintenance efforts in the future.

\section{Related Work}
\label{related}

There are some existing approach for detect redundancy in the code (i.e., code clone), such as \cite{baxterCloneDetectionUsing1998,kamiyaCCFinderMultilinguisticTokenbased2002}. Due to LLM's probabilistic model-based nature~\cite{bengioNeuralProbabilisticLanguage2003,shumailovCurseRecursionTraining2024}, LLM models typically do not generate textually identical code clones. Instead, they tend to produce more Type-4 clones~\cite{roySurveySoftwareClone2007}, redundant code that exhibits textual inconsistencies but shares similar semantic meaning. However, existing tools are not effective at detecting these Type-4 semantic clones in Agentic-PRs. Early approaches, such as \cite{sasakiFindingFileClones2010}, detect clones by comparing file hashes. This approach is too coarse because agents often add specific functions rather than duplicating entire files. Later approaches, such as \cite{kimFaCoYCodetocodeSearch2018}, use information retrieval techniques to find similar code, but they rely on keyword matching and often miss semantic similarities when the syntax is different.

\section{Threats to Validity}
\label{threats}

\textit{Internal Validity} 
We relied on PyRef to filter out refactorings (Move Method/Rename Method). 
If PyRef fails to detect a refactoring, we might incorrectly label valid code movement as redundancy. To mitigate this, we manually verified 10 sample of the results. Additionally, we only analyzed the crewAI repository for RQ1 due to computational costs. While this repository is highly active, it may not represent all coding styles and results might not generalize.

\textit{External Validity} We focused only on Python repositories. The behavior of AI agents may differ in other languages like Java or C++. Furthermore, our sentiment analysis model is trained on general English text. It may misinterpret technical discussions in code reviews as neutral or negative emotions.

\section{Conclusion}
\label{conclusion}
This study investigates the impact of AI coding agents in SE beyond simple pass rates. We combined a semantic redundancy analysis with a reviewer sentiment analysis.
We found that AI agents produce more redundant code (Type-4 clones) than human developers. However, human reviewers do not punish this behavior; instead, they show less negative sentiment towards AI-generated PRs. We conclude that the surface-level plausibility of AI code can mask poor design choices. This leads to hidden technical debt, where redundancy accumulates without being noticed. As the use of AI agents increases, developers need to maintain high review standards to ensure long-term code maintainability.

\begin{acks}
This work was partly developed through discussions held during the AI-Driven SE Summit 2025.
\end{acks}

\bibliographystyle{ACM-Reference-Format}
\bibliography{ref}

\end{document}